\documentstyle[prd,aps,preprint]{revtex}
\tightenlines
\input epsf.tex
\def\DESepsf(#1 width #2){\epsfxsize=#2 \epsfbox{#1}}
\begin{document}

\draft
%@twocolumnfalse\endcsname
\preprint{\hbox{CTP-TAMU-48-99}}
\title{Neutralino Proton Cross Sections In Supergravity Models} 

\author{ E. Accomando, R. Arnowitt, B. Dutta and Y. Santoso }

\address{
Center For Theoretical Physics, Department of Physics, Texas A$\&$M
University, College Station TX 77843-4242}
\date{December, 1999}
\maketitle
\begin{abstract}
The neutralino-proton cross section is examined for supergravity models with
R-parity invariance with universal and non-universal soft breaking. The region
of parameter space that dark matter detectors are currently (or will be shortly)
sensitive i.e. $(0.1-10)\times 10^{-6}$ pb, is examined. For universal soft
breaking (mSUGRA), detectors with sensitivity $\sigma_{\tilde{\chi}_{1}^{0}-p}
\geq 1 \times 10^{-6}$ pb will be able to sample parts of the parameter space for
$\tan \beta \stackrel{>}{\sim} 25$. Current relic density bounds restrict
$m_{\tilde{\chi}_{1}^{0}} \leq 120$ GeV for the maximum cross sections, which is
below where astronomical uncertainties about the Milky Way are relevant.
Nonuniversal soft breaking models can allow much larger cross sections and can
sample the parameter space for $\tan \beta \stackrel{>}{\sim} 4$. In such
models, $m_0$ can be quite large reducing the tension between proton decay
bounds and dark matter analysis. We note the existance of two new domains where
coannihilation effects can enter, i.e. for mSUGRA at large $\tan \beta$, and for
nonuniversal models with small $\tan \beta$.
\end{abstract}

\section{Introduction}
Supersymmetric models with R-parity invariance generally predict the existance
of dark matter relics from the Big Bang. Experimental bounds on exotic isotopes
strongly imply that the lightest supersymmetric particle (LSP), which is
absolutely stable by R-parity invariance, must be electrically neutral and
weakly interacting. The minimal supersymmetric model (MSSM) then has two
possible candidates, the lightest neutralino ($\tilde{\chi}^{0}_{1}$) and the
sneutrino ($\tilde{\nu}$). In gravity mediated supergravity (SUGRA) grand
unified models (GUTs)~\cite{chamseddine}, the allowed region in the supersymmetry (SUSY)
parameter space where the $\tilde{\nu}$ is the LSP is generally small. The
absence of the decay $Z \rightarrow \tilde{\nu} + \tilde{\nu}$ at LEP implies
$m_{\tilde{\nu}} \geq 45$ GeV and LEP bounds on the light Higgs is sufficient to
eliminate the $\tilde{\nu}$ as the LSP for the minimal SUGRA model
(mSUGRA)~\cite{arnowitt1}. 
If one further includes cosmological constraints, the sneutrino is
also excluded for the general MSSM~\cite{falk}. Thus for these models, the
$\tilde{\chi}^{0}_{1}$ is the unique candidate for cold dark matter (CDM). It is
an appealing feature then of SUGRA models that the predicted amount of relic
density of neutralino CDM is consistent with what is observed astronomically for
a significant part of the SUSY parameter space.

Since the initial observation that the $\tilde{\chi}^{0}_{1}$ represented a
possible CDM particle~\cite{goldberg} and the subsequent suggestion that local
$\tilde{\chi}^{0}_{1}$ in the Milky Way might be observed by terrestrial
detectors~\cite{goodman}, there has been a great deal of theoretical analysis and
experimental activity concerning the detection of local CDM particles. Recent theoretical
calculations in Refs. [6-25] have made use of a number of different SUSY models.
Thus Refs. [6-10] assume the MSSM model, and calculations in
Refs. [11-19] are performed using mSUGRA GUT models (with universal soft breaking
at the GUT scale $M_{G} \cong 2 \times 10^{16}$ GeV).
Refs.~\cite{berezinsky1,berezinsky2,bottino3} allow for
nonuniversal soft breaking in the Higgs sector and
Refs.~\cite{nath1,arnowitt4,arnowitt5} include 
also nonuniversal effects in the third generation. In addition, different
authors limit the parameter space differently.

The neutralino-nucleus scattering amplitude contains spin independent and spin
dependent parts. However for detectors with heavy nuclei, the spin independent
part dominates. In these, the neutron and proton scattering amplitudes are
approximately equal, which allows one to extract from the data the spin
independent neutralino-proton cross section $\sigma_{\tilde{\chi}^{0}_{1}-p}$.
The sensitivity of current experiments (e.g., DAMA, CDMS) is approximately
$(1-10) \times 10^{-6}$ pb for $\sigma_{\tilde{\chi}^{0}_{1}-p}$~\cite{bernabei}, and perhaps a
factor of 10 improvement may be expected in the near future. It is the purpose
of this paper to examine what part of the SUSY parameter space can be tested
with such a sensitivity. We do this by examining the maximum theoretical cross
section that can lie in the domain
\begin{equation}
0.1 \times 10^{-6}\ \rm{pb} \leq \sigma_{\tilde{\chi}^{0}_{1}-p} \leq 10 \times
10^{-6}\ \rm{pb}	\label{eq1}
\end{equation}
as one varies SUSY parameters (e.g., $\tan \beta$, $m_{\tilde{\chi}^{0}_{1}}$). Our
calculations are done within the framework of SUGRA GUT models with
non-universal soft breaking allowed in both the Higgs and third generation
squark and slepton sectors. (As discussed in Ref.~\cite{nath1} and will be seen
below, it is necessary to include both Higgs and third generation
nonuniversalities as the two can have constructive interference.) We also update
earlier analyses by including the latest LEP bounds on the light chargino
($\tilde{\chi}^{\pm}_{1}$) and light Higgs ($h$) mass ($m_{\tilde{\chi}^{\pm}_{1}}
> 94$ GeV, $m_{h} > 95$ GeV) and include the $b \rightarrow
s+ \gamma$ and Tevatron constraints. 

In calculating $\sigma_{\tilde{\chi}^{0}_{1}-p}$, we restrict the SUSY parameter
space to be consistent with the astronomical estimates of the amount of relic
CDM. This is conventionally measured by the quantity $\Omega_{\rm{CDM}} =
\rho_{\rm{CDM}}/\rho_{c}$ where $\rho_{\rm{CDM}}$ is the mean CDM mass density, and
$\rho_{c}=3 H_{0}^{2}/8 \pi G_{N}$ ($H_{0}$ = Hubble constant parameterized by
$H_{0}=(100\ \mbox{km s}^{-1}\ \mbox{Mpc}^{-1})h$, $G_{N}$ = Newton constant). Recent
measurements of $\Omega_{m} = \rho_{m}/\rho_{c}$ ($\rho_{m}$ is the matter
density), $H_{0}$, the Cosmic Microwave Background (CMB), supernovae data, etc., indicate that
$\Omega_{\rm{CDM}}$
is smaller then previously thought. We will see below that both the upper and
lower bounds on $\Omega_{\tilde{\chi}^{0}_{1}}h^{2}$ strongly affect the predicted
values of $\sigma_{\tilde{\chi}^{0}_{1}-p}$.

In our analysis below we use the one loop renormalization group equations
(RGE)~\cite{barger2} 
from $M_{G}$ to the $t$-quark mass $m_{t}=175$ GeV and impose the
radiative breaking constraint at the electroweak scale. We start with a set of parameters
at $M_{G}$, integrating out the heavy particles at each threshold, and iterate until a
consistent spectrum is obtained. One loop corrections are
included in diagonalizing the Higgs mass matrix, and L-R mixing is included in
the sfermion mass matrices so that large $\tan \beta$ may be treated.
Naturalness constraints, that the gluino mass obey $m_{\tilde{g}} \leq 1$ TeV,
the scalar mass $m_{0} \leq 1$ TeV and $|A_{0}/m_{0}| \leq 5$ are imposed.
Gaugino masses are assumed universal at $M_{G}$, and possible CP violating
phases are set to zero. Thus we do not treat here D-brane
models~\cite{brhlik2,accomando}
(which will be discussed in a subsequent paper). The SUSY mass spectrum is also
constrained so that coannihilation effects are negligible. (We find, in fact,
that this is a significant constraint with nonuniversal soft breaking even for
low $\tan \beta$.) We examine $\tan \beta$ in the range $2 \leq \tan \beta \leq
50$, and include leading order (LO) corrections to the $b \rightarrow s + \gamma$
decay and correct approximately for NLO effects~\cite{anlauf,baer3}. We require that
the theoretical branching ratio lie in the range $1.9 \times 10^{-4} \leq
\mbox{BR}(B \rightarrow X_{s} \gamma ) \leq 4.5 \times 10^{-4}$, and use one loop
corrections to the $b$-quark mass so that $m_{b}$ takes on its experimental
value $m_{b}(m_{b})=(4.1-4.5)$ GeV~\cite{PDG}. (See Appendix. The loop correction is significant for large
$\tan{\beta}$ and stems from the part of the Lagrangian given by $-\mu^{\ast}
\lambda_{b} \tilde{b}_{L} \tilde{b}_{R}^{\dag} H_{2}^{0 \dag} + h.c.$.) We do
not assume any particular GUT group constraints and do not impose $b-\tau$
Yukawa unification (since the latter is sensitive to possible unknown GUT
physics).

In Sec. 2 we discuss the range of the astrophysical parameters that enter into
the relic density analysis, and also the uncertainties of the quark content of
the proton which affect our calculation of $\sigma_{\tilde{\chi}^{0}_{1}-p}$. In
Sec. 3
we examine the mSUGRA model where it is seen that  $\sigma_{\tilde{\chi}^{0}_{1}-p}>1
\times 10^{-6}$ pb (the current experimental sensitivity) requires $\tan \beta$ to be quite large, though this is
somewhat relaxed in the domain $0.1 \times 10^{-6}\ \rm{pb} \leq \sigma_{\tilde{\chi}^{0}_{1}-p}
\leq 1 \times 10^{-6}\ \rm{pb}$. In Sec. 4 we discuss the nonuniversal models,
and see
that here Eq. (\ref{eq1}) can be satisfied for relatively small $\tan \beta$ and large
$m_{0}$, and also that $\sigma_{\tilde{\chi}^{0}_{1}-p}$ sustains for large
$m_{\tilde{\chi}^{0}_{1}}$. The SUSY mass spectrum expected for our domain of cross
sections is also examined. Conclusions are summarized in Sec. 5. A brief
qualitative discussion is also given there of the effect of these results on
proton decay since the above nonuniversal results appear to releave
some of the tension previously noted~\cite{arnowitt4} between
$\sigma_{\tilde{\chi}^{0}_{1}-p}$ in the range of Eq. (\ref{eq1}) and current Super Kamiokande
proton lifetime bounds~\cite{superK}.

\section{Astronomical and Quark Parameters}

The basic experimental quantity that controls the SUSY analysis of relic
density is $\Omega_{\tilde{\chi}^{0}_{1}}h^{2}$. Recent measurements at the Hubble
Space Telescope using a number of different techniques has led to a combined
average of~\cite{freedman}
\begin{equation}
H_{0} = (71 \pm 3 \pm 7)\ \rm{km\ s^{-1}\ Mpc^{-1}}\ .	\label{eq2}
\end{equation}
There is now sufficient data on the CMB
anisotropies to show that $\Omega_{tot} \simeq 1$ and $\Omega_{m}$ small is
strongly favored~\cite{dodelson}. Measurements on clusters of galaxies yield
$\Omega_{m} < 0.32 \pm 0.05$~\cite{mohr,dodelson}, and these results are consistent with
the supernovae data~\cite{perlmutter}. An analysis of combined data (excluding
microlensing) yields~\cite{lineweaver} $\Omega_{m} = 0.23 \pm 0.08$. In view of
possible systematic errors, we assume here $\Omega_{m} = 0.3 \pm 0.1$, and since
the baryonic content is $\Omega_{B} \cong 0.05$, we take 
\begin{equation}
\Omega_{\tilde{\chi}^{0}_{1}} = 0.25 \pm 0.10\ .	\label{eq3}
\end{equation}
Combining errors in quadrature then yields $\Omega_{\tilde{\chi}^{0}_{1}}h^{2} =
0.126 \pm 0.052$. In the following we will restrict the range of
$\Omega_{\tilde{\chi}^{0}_{1}}h^{2}$ by what is approximately 2 std. around the
mean: 
\begin{equation}
0.02 \leq \Omega_{\tilde{\chi}^{0}_{1}}h^{2} < 0.25\ .	\label{eq4}
\end{equation}
(As pointed out in Ref.~\cite{baer3}, the lower bound is the minimum amount of DM
to account for the rotation curves of spiral galaxies.) Future measurements by
the MAP and Planck sattelites will greatly reduce these errors.

The fundamental SUSY Lagrangian allows one to calculate the neutralino-quark
scattering amplitude. To obtain the $\tilde{\chi}^{0}_{1}-p$ cross section one
needs to know in addition, the quark content of the proton. In the notation of
Ref.~\cite{bottino2}, the two parameters that enter sensitively are
\begin{equation}
\hat{f} = \frac{\sigma_{\pi N}}{m_{p}}\ , 	\label{eq5}
\end{equation}
and 
\begin{equation}
f = \frac{\langle p | m_{s} \bar{s} s | p \rangle}{m_{p}}\ ,	\label{eq6} 
\end{equation}
where $\sigma_{\pi N}$ is the ($\pi-N$) $\sigma$-term and is given by 
\begin{equation}
\sigma_{\pi N} = \frac{1}{2} (m_{u} + m_{d}) \langle p | \bar{u} u + \bar{d} d | p
\rangle\ .	\label{eq7}
\end{equation}
$f$ can be written as~\cite{bottino2}
\begin{equation}
f=\frac{1}{2} r y \frac{\sigma_{\pi N}}{m_{p}}\ ,	\label{eq8}
\end{equation}
where
\begin{equation}
r = \frac{m_{s}}{\frac{1}{2}(m_{u}+m_{d})}\ ,	\label{eq9}
\end{equation}
and 
\begin{equation}
y = \frac{ \langle p | \bar{s} s | p \rangle }{\frac{1}{2} \langle p | \bar{u} u +
\bar{d} d | p \rangle } \equiv 1 - \frac{\sigma_{0}}{\sigma_{\pi N}}\ .	
\label{eq10}
\end{equation}
The quark mass ratios are fairly well known, and we use in the following $r=24.4
\pm 1.5$~\cite{leutwyler}. Recently, the uncertainties in $\sigma_{\pi N}$ and
$\sigma_{0}$ have been analyzed in Ref.~\cite{bottino2}. They find
\begin{equation}
40 \ \rm{MeV} \stackrel{<}{\sim} \sigma_{\pi N} \stackrel{<}{\sim} 65 \ \rm{MeV}, \quad 30
\ \rm{MeV} \stackrel{<}{\sim} \sigma_{0} \stackrel{<}{\sim} 40 \ \rm{MeV}\ . 
\label{eq11}
\end{equation}
In the following we will consider two possible choices for $\sigma_{\pi N}$ and
$\sigma_{0}$:
\begin{equation}
\begin{array}{lll} \mbox{Set 1:  } & \sigma_{\pi N}=40\ \rm{MeV},\
\sigma_{0}=30\ \rm{MeV};
& \hat{f}=0.0480,\ f=0.195\ . \\ \mbox{Set 2:  } & \sigma_{\pi N}=65\ \rm{MeV},\
\sigma_{0}=30\ \rm{MeV};
& \hat{f}=0.0693,\ f=0.455\ . 
\end{array}	\label{eq12}
\end{equation}
Set 1 corresponds approximately to the original analysis of Ref.~\cite{ellis2} (and
is the most conservative possibility) while Set 2 is similar to the Set 2 of
Ref.~\cite{bottino2}. (Set 3 of Ref.~\cite{bottino2} gives considerably larger cross
sections.) In the following, we will use mostly Set 2 in showing our result, but
we will exhibit the difference between Set 1 and Set 2 in one case to illustrate
some of the uncertainties that exist.

\section{The mSUGRA Model}

We begin our analysis by examining the minimal SUGRA model which depends on four
parameters and the sign of the Higgs mixing parameter $\mu$. A convenient choice of parameters is $m_{0}$
(the universal scalar mass at $M_{G}$), $m_{1/2}$ (the universal gaugino mass at
$M_{G}$), $A_{0}$ (the cubic soft breaking mass at $M_{G}$) and
$\tan{\beta}=\langle H_{2} \rangle / \langle H_{1} \rangle$ (where $\langle
H_{1,2} \rangle$ gives rise to (down, up) quark masses). It is convenient
sometimes to replace $m_{1/2}$ by the gluino mass $m_{\tilde{g}} \cong
(\alpha_{3}/\alpha_{G})m_{1/2}$ ($\alpha_{G} \cong 1/24$ is the GUT scale gauge
coupling constant) or $m_{\tilde{\chi}_{1}^{0}}$ which also scales with
$m_{1/2}$. Our sign convention on the $\mu$ parameter is defined by the quadratic term in
the superpotential
\begin{equation}
W^{(2)} = \mu H_{1} H_{2} = \mu (H_{1}^{0} H_{2}^{0} - H_{1}^{-} H_{2}^{+}).
\label{eq13}
\end{equation}
(With this convention, the $b \rightarrow s+\gamma$ constraint eliminates mostly $\mu > 0$.)

\begin{figure}[htb]
\bigskip
\bigskip
\bigskip
\centerline{ \DESepsf(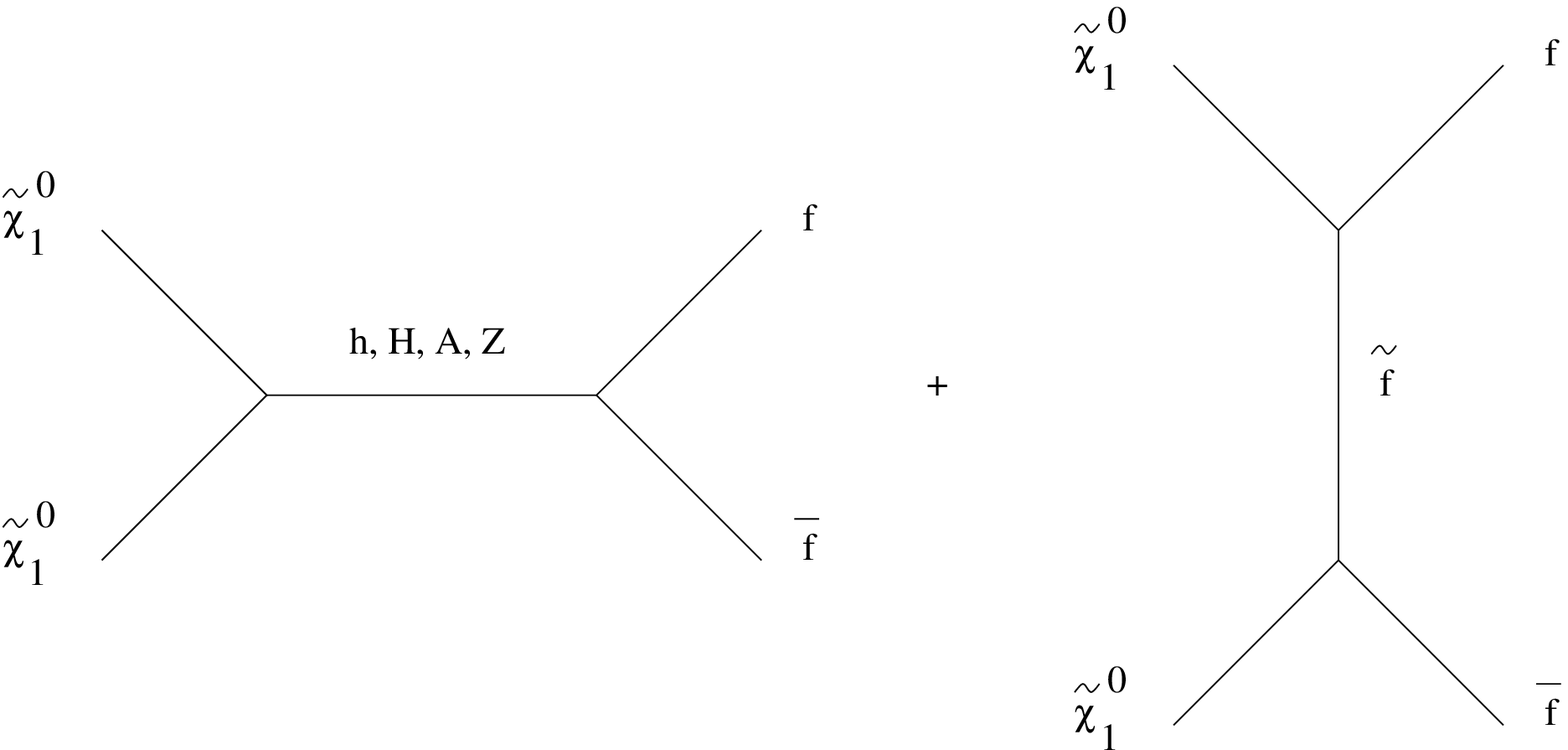 width 13 cm) }
\bigskip
\bigskip
\caption {\label{fig1} Diagrams for early universe annihilation of
$\tilde{\chi}_{1}^{0}$ through Higgs ($h$, $H$, $A$) and $Z$ poles and squark
and slepton ($\tilde{f}$) poles.}
\end{figure}

In calculating $\sigma_{\tilde{\chi}_{1}^{0}-p}$, one must impose the relic
density constraint of Eq. (\ref{eq4}). This is governed by the Boltzmann equation
describing $\tilde{\chi}_{1}^{0}$ annihilation in the early
universe~\cite{jungman}:
\begin{equation}
\frac{dn_{\tilde{\chi}_{1}^{0}}}{dt}+3 \frac{\dot{R}}{R}
n_{\tilde{\chi}_{1}^{0}} = \langle \sigma_{ann} v_{rel} \rangle
(n_{\tilde{\chi}_{1}^{0}}-n_{eq})	\label{eq14}
\end{equation}
where $n_{\tilde{\chi}_{1}^{0}}$ is the number density of
$\tilde{\chi}_{1}^{0}$, $n_{eq}$ its equilibrium value, $\sigma_{ann}$ is the
annihilation cross section, $v_{rel}$ the relative velocity, and $\langle \ 
\rangle$ means thermal average. The diagrams governing $\sigma_{ann}$ are shown in
Fig.~\ref{fig1}. The final relic density is given by
\begin{equation}
\Omega_{\tilde{\chi}_{1}^{0}}h^{2} = 2.48 \times 10^{-11} \left(
\frac{T_{\tilde{\chi}_{1}^{0}}}{T_{\gamma}} \right)^{3} \left(
\frac{T_{\gamma}}{2.73} \right)^{3} \frac{N_{f}^{1/2}}{\int_{0}^{x_{f}} dx \langle
\sigma_{ann} v_{rel} \rangle }		\label{eq15}
\end{equation}
where $x_{f} = kT_{f}/m_{\tilde{\chi}_{1}^{0}} \simeq 1/20 $, $T_{f}$ is the
freezeout temperature, $N_{f}$ is the number of degrees of freedom at freezeout,
and $(T_{\tilde{\chi}_{1}^{0}}/T_{\gamma})^{3}$ is the reheating factor.

The relic density decreases with increasing annihilation cross section, and in
order to understand some of the results obtained below, we first discuss which
parameters control $\sigma_{ann}$. From Fig.~\ref{fig1} one expects $\sigma_{ann}$ to
fall with increasing $m_{\tilde{\chi}_{1}^{0}}$ and also increasing $m_{0}$
(since $m_{\tilde{f}}^{2}$ increases with $m_{0}^{2}$). However, if $2
m_{\tilde{\chi}_{1}^{0}}$ is near $m_{h}$, $m_{H}$ or $m_{A}$ (but lies below),
the s-channel pole gives rise to a large amount of annihilation (which can
reduce $\Omega_{\tilde{\chi}_{1}^{0}}h^{2}$ below the allowed minimum) and due
to the thermal averaging, this effect can be significant when $2
m_{\tilde{\chi}_{1}^{0}}$ is less than the Higgs mass and within five times the 
Higgs width of the Higgs mass~\cite{greist,arnowitt6}. The LEP data, has eliminated most of this effect for the light
Higgs. However, we will see that since $H$ and $A$ become light at large
$\tan{\beta}$, effects of this type become significant in that regime. Further,
if one of the sleptons or squarks becomes light i.e. $\simeq 100$ GeV, the
t-channel annihilation will drive  $\Omega_{\tilde{\chi}_{1}^{0}}h^{2}$ 
down~\cite{roszkowski}. This effect can again become significant at large $\tan{\beta}$ where
large L-R mixing in the sfermion mass matrices reduces $m_{\tilde{f}}^{2}$.

\begin{figure}[htb]
\centerline{ \DESepsf(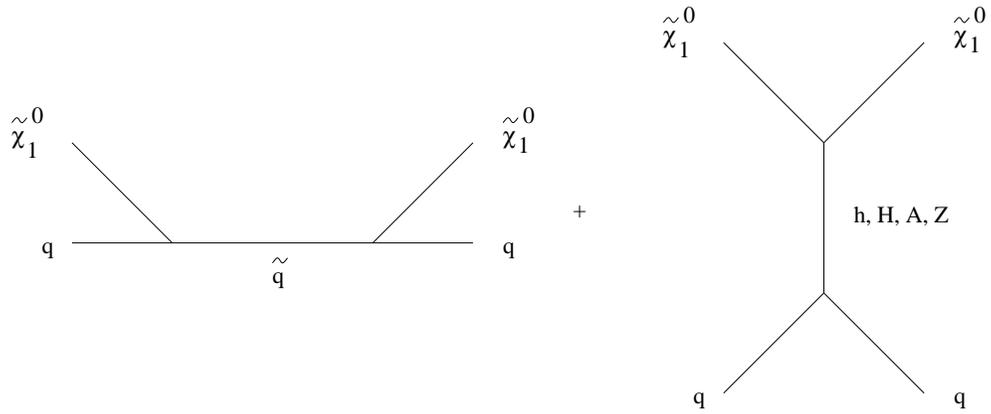 width 13 cm) }
\bigskip
\bigskip
\caption {\label{fig2} Diagrams for neutralino quark scattering.}
\end{figure}

We turn next to $\sigma_{\tilde{\chi}_{1}^{0}-p}$, which is governed by the
diagrams of Fig.~\ref{fig2}. We see here that the cross section can become large for light (first generation
squarks) and light Higgs bosons. These regions of parameter space are just the
ones that reduce $\Omega_{\tilde{\chi}_{1}^{0}}h^{2}$, and so there can be a
bound produced on $\sigma_{\tilde{\chi}_{1}^{0}-p}$ so that
$\Omega_{\tilde{\chi}_{1}^{0}}h^{2}$ does not fall below its minimum.

In order now to see the sensitivity of current detectors to mSUGRA we plot in
Fig.~\ref{fig3} (for Set 2 parameters of Eq. (\ref{eq12})), the maximum value of
$\sigma_{\tilde{\chi}_{1}^{0}-p}$ for $\tan{\beta}=$20, 30, 40 and 50 as a function of
$m_{\tilde{\chi}_{1}^{0}}$ (obtained by allowing  all other parameters to vary
subject to the constraints listed in Secs. 1 and 2). We see the expected fall off 
with increasing $m_{\tilde{\chi}_{1}^{0}}$. The
current DAMA experiment is thus sensitive to mSUGRA for $\tan{\beta}
\stackrel{>}{\sim} 25$ (i.e. $\sigma_{\tilde{\chi}_{1}^{0}-p} \stackrel{>}{\sim}
1.0 \times 10^{-6}$ pb). We note that the fall off is less severe for $\tan{\beta}=50$,
since at this high value of $\tan{\beta}$ the $H$ and $A$ Higgs become relatively light enhancing the
$\tilde{\chi}_{1}^{0}-p$ cross section. This can be seen in Fig.~\ref{fig4} where we have plotted
$m_{H}$ for $\tan{\beta}=30$ and $\tan{\beta}=50$. We note also the importance of
including the loop corrections to $m_{b}$ for large $\tan{\beta}$ (e.g. $\tan{\beta}=50$)
to obtain the correct results here. 

Fig.~\ref{fig5} shows the sensitivity of the calculations to the choice of
particle physics parameters. Set 1 gives cross sections about a factor of 2
smaller than Set 2. In the following, we will use Set 2 in all our analysis.

Fig.~\ref{fig6} shows $\Omega_{\tilde{\chi}_{1}^{0}}h^{2}$ vs $m_{\tilde{\chi}_{1}^{0}}$ for
$\tan{\beta}=30$. We see, as expected, $\Omega_{\tilde{\chi}_{1}^{0}}h^{2}$ is
an increasing function of $m_{\tilde{\chi}_{1}^{0}}$ (since $\sigma_{ann}$ is a
decreasing function). The upper bound on $\Omega_{\tilde{\chi}_{1}^{0}}h^{2}$
then implies an upper bound of the neutralino mass of $m_{\tilde{\chi}_{1}^{0}}
\cong 120$ GeV (i.e. $m_{\tilde{g}} \stackrel{<}{\sim} 900$ GeV) as has been discussed
previously [12-14]. Note that one can obtain cross sections within the DAMA
sensitivity range without going to the edges of the parameter space in
$\Omega_{\tilde{\chi}_{1}^{0}}h^{2}$. Also, these cross sections all fall below
the current experimental sensitivity before the uncertainties in the Milky Way
astronomical parameters discussed in~\cite{brhlik1,belli} become important.

Figs. (7-9) exhibit the particle spectrum expected for the example of
$\tan{\beta}=30$, when $\sigma_{\tilde{\chi}_{1}^{0}-p}$ takes on its maximum
value. Fig.~\ref{fig7} shows that the $d$-squark is quite heavy ($m_{0}$ is large). This
arises from our constraint $m_{\tilde{\tau}_{R}} - m_{\tilde{\chi}_{1}^{0}} \geq
25$ GeV to prevent coannihilation effects from occurring. Thus the large
$\tan{\beta}$ being considered here reduces $m_{\tilde{\tau}_{R}}$ (due to L-R
mixing) and $m_{0}$ must be increased to prevent it from becoming degenerate
with the $\tilde{\chi}_{1}^{0}$. These coannihilation effects are thus different
from the ones that can occur at low $\tan{\beta}$~\cite{berezinsky1}, since they occur at
low $m_{\tilde{\chi}_{1}^{0}}$ where $\sigma_{\tilde{\chi}_{1}^{0}-p}$ is large
enough to fall within the range of Eq. (\ref{eq1}). (They will be
discussed elsewhere.) Fig.~\ref{fig8} shows the light Higgs mass, which is relatively heavy
due to the fact that $m_{0}$ is large. Fig.~\ref{fig9} shows $m_{\tilde{\chi}_{1}^{\pm}}$ vs
$m_{\tilde{\chi}_{1}^{0}}$ for $\tan{\beta}=30$. One sees that scaling is 
obeyed~\cite{arnowitt7}, i.e. $m_{\tilde{\chi}_{1}^{\pm}} \cong 2 m_{\tilde{\chi}_{1}^{0}}$ since
$\mu$ is relatively large ($\mu^{2}/M_{Z}^{2} \gg 1$).

\section{Nonuniversal Models}

Nonuniversal soft breaking can arise in SUGRA models if in the Kahler potential, the
interactions between the fields of the hidden sector (that break supersymmetry) and
the physical sector are not universal. Nonuniversalities allow for a remarkable
increase in the neutralino-proton cross section.
 
In order to suppress flavor changing neutral currents (FCNC), we will assume the
first two generations of squarks and sleptons are universal at $M_{G}$ (with soft
breaking mass $m_{0}$) but allow for nonuniversalities in the Higgs and third
generation. 
Thus we parameterize the soft breaking mass at $M_{G}$ as follows:
\begin{eqnarray}
m_{H_{1}}^{\ 2}=m_{0}^{2}(1+\delta_{1}); \quad m_{H_{2}}^{\ 2}=m_{0}^{2}(1+ \delta_{2});
	\label{eq16}  \\
m_{q_{L}}^{\ 2}=m_{0}^{2}(1+\delta_{3}); \quad m_{u_{R}}^{\ 2}=m_{0}^{2}(1+\delta_{4});
\quad m_{e_{R}}^{\ 2}=m_{0}^{2}(1+\delta_{5}); 	\label{eq17}	\\
m_{d_{L}}^{\ 2}=m_{0}^{2}(1+\delta_{6}); \quad m_{l_{L}}^{\
2}=m_{0}^{2}(1+\delta_{7});	\label{eq18}
\end{eqnarray}
where $m_{0}$ is the universal mass of the first two generations,
$q_{L}=(\tilde{t}_{L}, \tilde{b}_{L})$; $l_{L}=(\tilde{\nu}_{L},
\tilde{\tau}_{L})$; $u_{R}=\tilde{t}_{R}$; $e_{R}=\tilde{\tau}_{R}$ etc. We take
here the bounds 
\begin{equation}
-1 \leq \delta_{i} \leq 1	\label{eq19}
\end{equation}
An alternate way of satisfying the FCNC constraint is to make the first two
generations very heavy, and only the third generation light. This is essentially
included in the above parameterization by making $m_{0}$ large, and taking the
$\delta_{i}$ sufficiently close to -1.

One of the important parameters effected by the nonuniversal soft breaking
masses is $\mu^{2}$, which is determined by the radiative breaking condition.
While the RGE must be solved numerically, an analytic expression can be obtained
for low and intermediate $\tan{\beta}$~\cite{nath1}:
\begin{eqnarray}
\mu^{2} & = & \frac{t^{2}}{t^{2}-1} \left[ \left\{ \frac{1-3 D_{0}}{2} +
\frac{1}{t^{2}} \right\} + \left\{
\frac{1-D_{0}}{2}(\delta_{3}+\delta_{4})-\frac{1+D_{0}}{2} \delta_{2} +
\frac{\delta_{1}}{t^{2}} \right\} \right] m_{0}^{2} \nonumber \\ & & + \mbox{universal parts +
loop corrections}	\label{eq20}
\end{eqnarray}
where $t \equiv \tan{\beta}$ and 
\begin{equation}
D_{0} \cong 1 - (\frac{m_{t}}{200 \sin{\beta}})^{2} .	\label{eq21}
\end{equation}
A similar expression holds for large $\tan{\beta}$ in the $SO(10)$ limit so
Eq. (\ref{eq20}) gives a qualitative picture of the effects of nonuniversalities in
general (a result borne out from detailed numerical calculations).

We see first that in general $D_{0}$ is small, i.e. for $m_{t}=175$ GeV, $D_{0}
\leq 0.2$, and hence
the squark nonuniversality, $\delta_{3}$ and $\delta_{4}$, produce comparable
size effects as the Higgs nonuniversalities $\delta_{1}$ and $\delta_{2}$, so
that both must be included for a full treatment~\cite{nath1}. Second, one can choose
the signs of $\delta_{i}$ such that either $\mu^{2}$ is reduced or $\mu^{2}$ is
increased. The significance of this is that in general, the
$\tilde{\chi}_{1}^{0}$ is a mixture of Higgsino and gaugino pieces
\begin{equation}
\tilde{\chi}_{1}^{0} = \alpha \tilde{W}_{3} + \beta \tilde{B} + \gamma
\tilde{H}_{1} + \delta \tilde{H}_{2}	\label{eq22}
\end{equation}
Now the spin independent part of $\tilde{\chi}_{1}^{0}-q$ scattering depends on
interference between the gaugino and Higgsino parts of
$\tilde{\chi}_{1}^{0}$~\cite{ellis2} 
(it would vanish for pure
gaugino or pure Higgsino) and this interference increases if $\mu^{2}$ is
decreased (increasing $\sigma_{\tilde{\chi}_{1}^{0}-q}$) and decreases if
$\mu^{2}$ is increased (decreasing
$\sigma_{\tilde{\chi}_{1}^{0}-q}$)~\cite{arnowitt3}.
Thus there are regions in the parameter space of nonuniversal models where
$\sigma_{\tilde{\chi}_{1}^{0}-p}$ is significantly increased compared to the
universal case.

The above effect can be seen in Fig.~\ref{fig10} where the maximum
$\sigma_{\tilde{\chi}_{1}^{0}-p}$ are plotted for $\tan{\beta}=7$ for the
nonuniversal and universal cases. We see that nonuniversalities can increase
$\sigma_{\tilde{\chi}_{1}^{0}-p}$ by a factor of $\simeq 10$. Fig.~\ref{fig11} plots the
nonuniversal curves for $\tan{\beta}=3,5$, and 7. One sees here that with
nonuniversal soft breaking, the current DAMA sensitivity requires $\tan{\beta}
\stackrel{>}{\sim} 4$ (compared to $\tan{\beta} \stackrel{>}{\sim} 25$  in the universal case). For
larger $\tan{\beta}$ one can get very large nonuniversal cross sections.
Fig.~\ref{fig12}
shows the maximum $\sigma_{\tilde{\chi}_{1}^{0}-p}$ for $\tan{\beta}=15$, which
already lies in the region excluded by CDMS and DAMA.

For GUT groups containing an $SU(5)$ subgroup (such as $SU(5)$, $SO(10)$,
$SU(6)$ etc.) with matter in the usual $10+\bar{5}$ representations, the
$\delta_{i}$ of Eqs. (17,18) obey 
\begin{equation}
\delta_{3} = \delta_{4} = \delta_{5} \equiv \delta_{10}; \quad \delta_{6} = \delta_{7}
\equiv \delta_{\bar{5}}		\label{eq23}
\end{equation}
We consider this case in more detail (where it is assumed that the gauge group
breaks to the Standard Model at $M_{G}$). Fig.~\ref{fig13} shows
$\Omega_{\tilde{\chi}_{1}^{0}}h^{2}$ when $\sigma_{\tilde{\chi}_{1}^{0}-p}$
takes on its maximum value for the characteristic example of $\tan{\beta}=7$.
One sees that $\Omega_{\tilde{\chi}_{1}^{0}}h^{2}$ is generally small since one
has $\delta_{10} < 0$ to obtain the maximum $\sigma_{\tilde{\chi}_{1}^{0}-p}$
(reducing $\mu^{2}$ in Eq. (\ref{eq20}) and hence increasing the cross section). This
however reduces $m_{\tilde{\tau}_{R}}$ (from Eq. (\ref{eq17})) increasing the
annihilation rate as in the discussion of Fig.~\ref{fig1}. (If the $SU(5)$-type
constraint were relaxed and $\delta_{5}$ left arbitrary,
$\Omega_{\tilde{\chi}_{1}^{0}}h^{2}$ could be increased. For example,
$\delta_{5}=0$ produces $\approx 50 \% $ increase in
$\Omega_{\tilde{\chi}_{1}^{0}}h^{2}$.) The further fall off of
$\Omega_{\tilde{\chi}_{1}^{0}}h^{2}$ for $m_{\tilde{\chi}_{1}^{0}}
\stackrel{>}{\sim} 110$ GeV arises from the fact that $m_{H} \simeq 300$ GeV,
and the nearness of the $m_{H}$ s-channel pole of Fig.~\ref{fig1} increases the early universe annihilation. This
can be seen explicitly in Fig.~\ref{fig14} where $2 m_{\tilde{\chi}_{1}^{0}}$ is
close to $m_{H}$
when $m_{\tilde{\chi}_{1}^{0}} \stackrel{>}{\sim} 110$ GeV. Fig.~\ref{fig15} shows that the light Higgs for this
case is quite light lying just above the LEP2 bounds. Particularly interesting
is that the first two generations of squarks, however, are relatively heavy. 
This is shown in Fig.~\ref{fig16} for the $d$-squark. The reason for this can be seen from
Eq. (\ref{eq20}) where since $\delta_{3}=\delta_{4}=\delta_{10}<0$ (to lower $\mu^{2}$
and hence increase $\sigma_{\tilde{\chi}_{1}^{0}-p}$) the nonuniversal terms
produce a net negative $m_{0}^{2}$ contribution to $\mu^{2}$, the lowering of
$\mu^{2}$ being enhanced, then, the larger $m_{0}$ is. Thus it is possible to
get heavy squarks in the first two generations at low $\tan{\beta}$, which may
have implications with respect to proton decay as discussed in the next section.

\section{Conclusions}

If the dark matter of the Milky Way is indeed mainly neutralinos, then current
detectors are now sensitive to interesting parts of the SUSY parameter space.
Thus either discovery (or lack of discovery) will determine (or eliminate) parts
of the parameter space, and this analysis is complementary to what one may learn
from accelerator experiments.

To examine what parts of the parameter can be tested with current detectors or
in the near future, we have considered $\sigma_{\tilde{\chi}_{1}^{0}-p}$, the
$\tilde{\chi}_{1}^{0}-p$ cross section, in the range $0.1 \times 10^{-6} \ \rm{pb} \leq
\sigma_{\tilde{\chi}_{1}^{0}-p} \leq 10 \times 10^{-6}$ pb, and have plotted the
maximum theoretical cross section for different SUGRA models. There is a major
difference between the universal and nonuniversal soft breaking models. Thus the
current DAMA experiment (with sensitivity of $\sigma_{\tilde{\chi}_{1}^{0}-p}
\stackrel{>}{\sim} 1 \times 10^{-6}$ pb) is sensitive to $\tan{\beta}
\stackrel{>}{\sim} 25$ for universal soft breaking (Fig.~\ref{fig3}) while it is sensitive
to $\tan{\beta} \stackrel{>}{\sim} 4$ for the nonuniversal model (Fig.~\ref{fig11}). Thus
while dark matter cross sections increase with $\tan{\beta}$ and hence detectors
are more sensitive at higher $\tan{\beta}$, it is possible for current detectors
to probe part of the low $\tan{\beta}$ parameter space for the nonuniversal
models.

For the mSUGRA model, we find that $\Omega_{\tilde{\chi}_{1}^{0}}h^{2}$
monotonically increases with $m_{\tilde{\chi}_{1}^{0}}$ from the minimum to the
maximum bounds of Eq. (\ref{eq4}) (Fig.~\ref{fig6}), leading to the upper bound
$m_{\tilde{\chi}_{1}^{0}} \leq 120$ GeV ($m_{\tilde{g}} \stackrel{<}{\sim} 900$
GeV) which is below
where astronomical uncertainties about the Milky Way~\cite{brhlik1,belli} become significant. In
general $\mu^{2}$ is large (i.e. $\mu^{2}/M_{Z}^{2} \gg 1$) leading to the usual
gaugino scaling relations e.g. Fig.~\ref{fig9}, and the Higgs mass is relatively heavy
(Fig.~\ref{fig8}). At the very largest $\tan{\beta}$, e.g. $\tan{\beta}=50$, the loop
corrections to $\lambda_{b}$ at the electroweak scale become very large (see
Appendix), requiring that $\lambda_{b}$, the $b$-Yukawa coupling, be adjusted so that one
obtains the experimental $b$-quark mass~\cite{PDG}.

For the  nonuniversal model, significantly increased $\tilde{\chi}_{1}^{0}-p$
cross sections can be obtained by choosing $\delta_{3,4} < 0$ and $\delta_{2}>0$
in Eq. (\ref{eq20}). This reduces $\mu^{2}$, increasing the Higgsino content of the
$\tilde{\chi}_{1}^{0}$, and hence increasing the Higgsino-gaugino interference
which enters in $\sigma_{\tilde{\chi}_{1}^{0}-p}$. (In the $SU(5)$-like models,
this generally leads to a light $\tilde{\tau}_{R}$ and hence a relatively low
$\Omega_{\tilde{\chi}_{1}^{0}}h^{2}$ (Fig.~\ref{fig13})). In this case the maximum cross
sections arise with $\mu^{2}$ relatively small, and so scaling no longer holds
accurately, and the light Higgs lies close to the LEP2 bounds (Fig.~\ref{fig15}).

While coannihilation effects have not been treated in this analysis, we have
noted two regions where such effects can occur. In mSUGRA models, due to the
fact that $\tan{\beta}$ must be large to obtain
$\sigma_{\tilde{\chi}_{1}^{0}-p}$ in the range of Eq. (\ref{eq1}), L-R mixing reduces
$m_{\tilde{\tau}_{R}}$ making the $\tilde{\tau}_{R}$ near degenerate with the
$\tilde{\chi}_{1}^{0}$. In the nonuniversal case, where $\tan{\beta}$ is small
or moderate, large $\sigma_{\tilde{\chi}_{1}^{0}-p}$ are obtained by lowering
$\mu^{2}$ which makes the $\tilde{\chi}_{1}^{\pm}$ nearly degenerate with the
$\tilde{\chi}_{1}^{0}$. Both these domains of coannihilation are different that
previously treated~\cite{ellis1}, and they inhabit regions of parameter space with
$\sigma_{\tilde{\chi}_{1}^{0}-p}$ within the reach of current detectors. We have
prevented coannihilation here from becoming significant by imposing the
constraints $m_{\tilde{\tau}_{R}}-m_{\tilde{\chi}_{1}^{0}}$,
$m_{\tilde{\chi}_{1}^{\pm}}-m_{\tilde{\chi}_{1}^{0}} \geq 25$ GeV. Further study
is required to see what occurs when these constraints are removed.

It has for sometime been realized that tension exist in GUT theories that
simultaneously allow for dark matter and proton decay~\cite{arnowitt4,goto}. Thus for
$SU(5)$-type models, minimal SUGRA GUT proton decay proceeds through the
$\tilde{H}_{3}$, the superheavy Higgsino color triplet components of the Higgs
$5$ and $\bar{5}$ representations. The basic diagram is shown in Fig.~\ref{fig17}, showing
that the decay rate scales approximately by 
\begin{equation}
\Gamma(p \rightarrow \bar{\nu} K) \sim \frac{1}{M_{3}^{2}} \left(
\frac{m_{\tilde{\chi}_{1}^{\pm}}}{m_{\tilde{q}}^{2}} \frac{1}{\sin{\beta}
\cos{\beta}} \right)^{2}	\label{eq24}
\end{equation}
where $M_{3}=O(M_{G})$ is the $\tilde{H}_{3}$ mass. In mSUGRA models, scaling is
generally a good approximation and $m_{\tilde{\chi}_{1}^{\pm}} \cong 2
m_{\tilde{\chi}_{1}^{0}}$. Hence proton stability requires small
$m_{\tilde{\chi}_{1}^{0}}$, large $m_{\tilde{q}}$ and small $\tan{\beta}$. We
have seen, however, that if dark matter exists with the sensitivity of the
current DAMA experiment, while moderately heavy squark masses could exist in
mSUGRA (Fig.~\ref{fig7}), $\tan{\beta}$ would have to be quite large i.e. $\tan{\beta}
\stackrel{>}{\sim} 25$, which would be sufficient to violate the current Super
Kamiokande bounds on the proton lifetime~\cite{superK}. However, this tension is
releaved for the nonuniversal SUGRA GUT models. Thus we saw in these cases, one
could have a $\sigma_{\tilde{\chi}_{1}^{0}-p}$ in the range of the DAMA
experiment for small $\tan{\beta}$, i.e. $\tan{\beta} \stackrel{>}{\sim} 4$, and
further such large cross sections also implied large squark masses, Fig.~\ref{fig16}. This
would be expected to remove any disagreement between a large
$\sigma_{\tilde{\chi}_{1}^{0}-p}$ and a small proton decay rate.

Finally we mention that in this paper we have plotted the maximum
$\tilde{\chi}_{1}^{0}-p$ cross sections for each $\tan{\beta}$ and
$m_{\tilde{\chi}_{1}^{0}}$. Of course nature may not chose SUSY parameters such
that $\sigma_{\tilde{\chi}_{1}^{0}-p}$ takes on its maximum value. However, by
looking at the maximum $\sigma_{\tilde{\chi}_{1}^{0}-p}$ we are able to see in a
given model whether detection of dark matter at current detector sensitivities is
consistent with the predictions of the theoretical model.

\section{Acknowledgments}

This work was supported in part by National Science Foundation Grant No.
PHY-9722090.  

\section{Appendix}
The $b$-quark coupling to the down type Higgs field which gives rise to tree
level bottom mass is described by 
\begin{equation}
{ L_{bbH}}=\lambda_b \bar b_Lb_RH^0_1+h.c..	\label{eq25}
\end{equation}
There also exists a term in the Lagrangian where the bottom squarks  are
coupled to the up type neutral Higgs ($H_2^0$) and is given by:
\begin{equation} 
L_{\tilde b\tilde bH}=-\lambda_b \mu^*\tilde
b_L\tilde b_R^\dag{H^0_2}^\dag+h.c..
\end{equation} 
The above interaction can give rise to a
one loop  contribution to the tree level bottom mass~\cite{rattazi}. We do the
analysis in the mass insertion approximation which produces errors of less than
10$\%$ in $m_{b}$ for the relevant parts of the parameter space. 
The loop diagram arising from the above interaction, shown in
Fig. 18a, involves gluino, squark fields, $\alpha_s$ and $\tan\beta$ and hence can be
large for large $\tan\beta$. There  also exists another one loop
contribution which  involves the stop quarks and the chargino. This loop, shown
in Fig. 18b, depends on  $\lambda_t^2$ and contributes less than the gluino loop.
The net $b$-quark mass generated from the above contributions is
$m_b+\delta m_b$, where
\begin{equation}
\delta m_b=\lambda_b v_1 K \tan{\beta}; \quad \quad v_{1}=\langle H_{1}^{0}
\rangle
\end{equation}
\begin{eqnarray} 
K \simeq
-{2\alpha_{s}\over(3 \pi)} m_{\tilde g} \mu 
G (m_{\tilde{b}_L}^2, m_{\tilde{b}_R}^2, m_{\tilde g}^2)-
        {\lambda_t^2 \over (4\pi)^2} A_t \mu G(m_{\tilde{t}_L}^{2},
    m_{\tilde{t}_R}^{2},\mu^2) 
\end{eqnarray} 
where 
\begin{equation}
G(a,b, c)=\frac{a b \mbox{ Log}[\frac{a}{b}]+ b c \mbox{ Log}[\frac{b}{c}]+ a c
\mbox{ Log}[\frac{c}{a}]}{(a-b)(b-c)(a-c)} ,
\end{equation}
$m_{\tilde g}$ is the
gluino mass, $m_{\tilde{b}_{L,R}}$ are the left and right handed sbottom masses
and  $m_{\tilde{t}_{L,R}}$ are the left and right handed stop masses.

The correction $K$ is evaluated at the electroweak scale which we take here to
$m_{t}$ (the endpoint of running the RGE down from $M_{G}$). Using the RGE for
$\lambda_{b}$, we then determine $\lambda_{b}(m_{t})$ so that the total
$b$-quark mass, $m_{b}=\lambda_b v_1 + \delta m_{b}$, agrees with the
experimental value of $m_b(m_b)$~\cite{superK} at the $b$ scale. This produces a
significant change in $\lambda_b$ for large $\tan{\beta}$.

\bigskip
\bigskip
\bigskip

\bigskip
\bigskip
\bigskip

\begin{figure}[htb]
\centerline{ \DESepsf(aads20304050.epsf width 12 cm) }
\bigskip
\bigskip
\caption {\label{fig3} The maximum value of $\sigma_{\tilde{\chi}_{1}^{0}-p}$
vs. $m_{\tilde{\chi}_{1}^{0}}$ for $\tan{\beta}=20$, 30, 40, and 50 for Set 2
parameters of Eq. (\ref{eq12}).}
\end{figure}
\bigskip
\bigskip
\bigskip
\begin{figure}[htb]
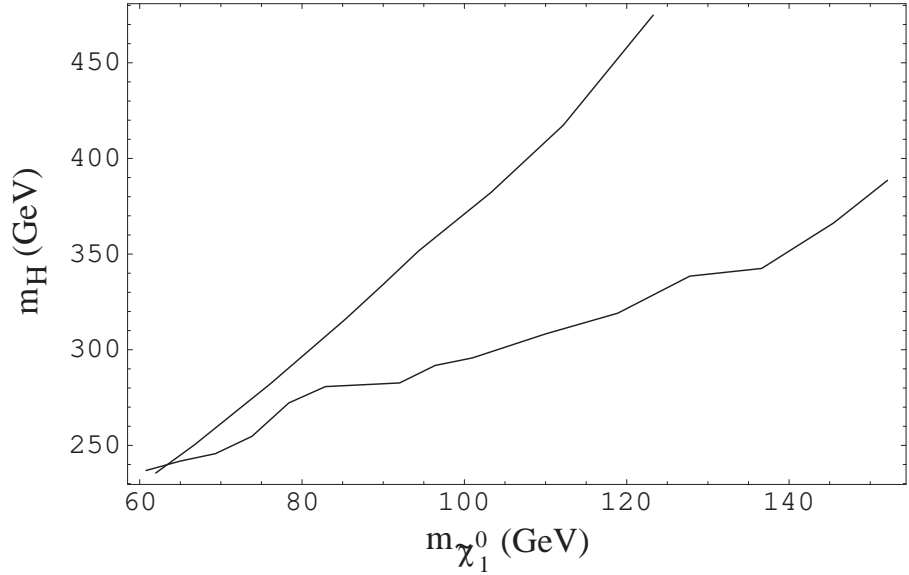

\centerline{ \DESepsf(aads3050Higgs.epsf width 12 cm) }
\bigskip
\bigskip
\caption {\label{fig4} $m_{H}$ vs. $m_{\tilde{\chi}_{1}^{0}}$ when
$\sigma_{\tilde{\chi}_{1}^{0}-p}$ takes on its maximum value. Top curve is for
$\tan{\beta}=30$, bottom curve for $\tan{\beta}=50$.}
\end{figure}

\newpage
\begin{figure}[htb]
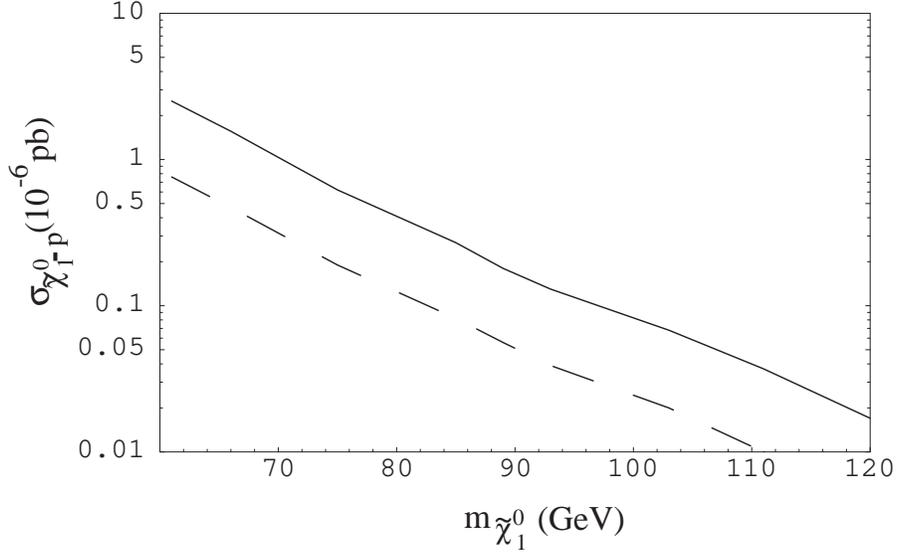

\centerline{ \DESepsf(aads3012.epsf width 12 cm) }
\bigskip
\bigskip
\caption {\label{fig5} Maximum $\sigma_{\tilde{\chi}_{1}^{0}-p}$
vs. $m_{\tilde{\chi}_{1}^{0}}$ for $\tan{\beta}=30$ for Set 2 parameters
(solid), and Set 1 parameters (dashed). See Eq. (\ref{eq12}).}
\end{figure}
\bigskip
\bigskip
\bigskip
\begin{figure}[htb]
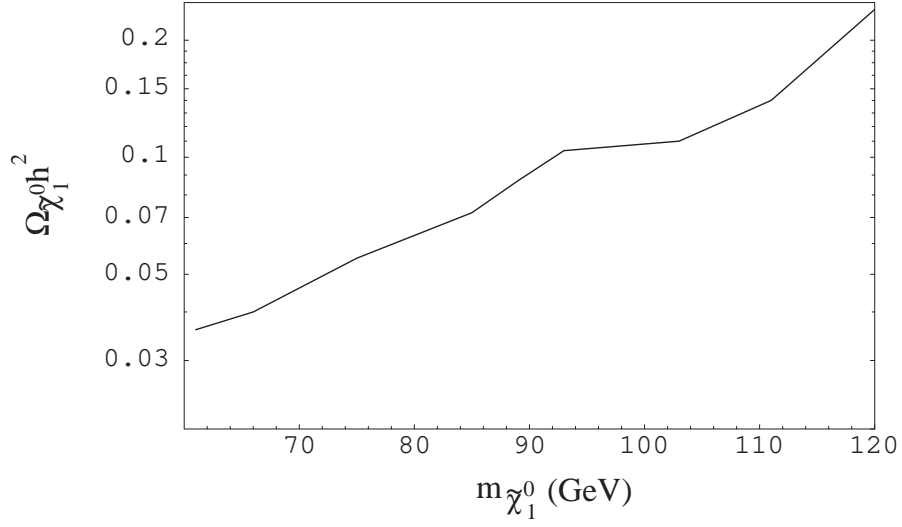

\centerline{ \DESepsf(aads30om.epsf width 12 cm) }
\bigskip
\bigskip
\caption {\label{fig6} $\Omega_{\tilde{\chi}_{1}^{0}}h^{2}$
vs. $m_{\tilde{\chi}_{1}^{0}}$ for $\tan{\beta}=30$.}
\end{figure}
\bigskip
\bigskip
\bigskip

\begin{figure}[htb]
\centerline{ \DESepsf(aads30dsqrk.epsf width 12 cm) }
\bigskip
\bigskip
\caption {\label{fig7} $m_{\tilde{d}}$
vs. $m_{\tilde{\chi}_{1}^{0}}$ for $\tan{\beta}=30$ when
$\sigma_{\tilde{\chi}_{1}^{0}-p}$ takes on its maximum value.}
\end{figure}
\bigskip
\bigskip
\bigskip
\begin{figure}[htb]
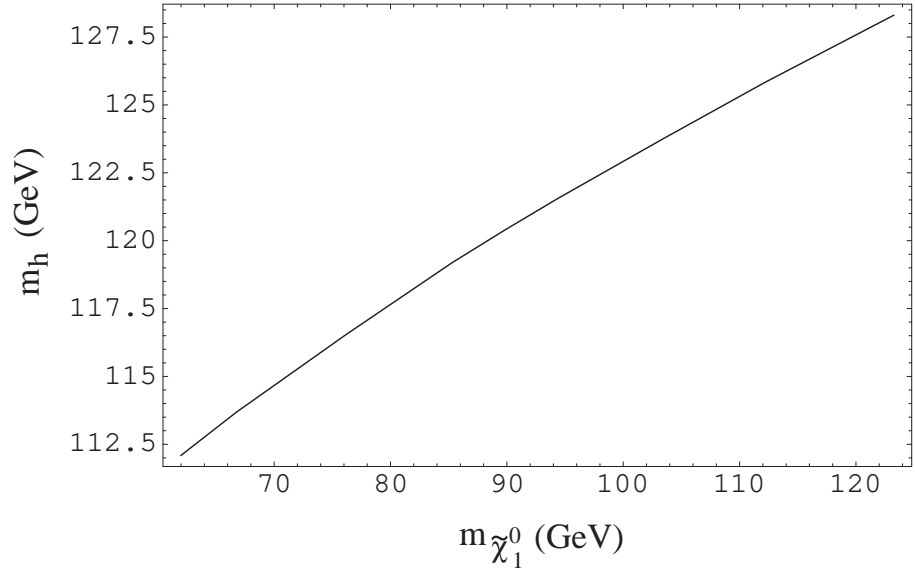

\centerline{ \DESepsf(aads30higgs.epsf width 12 cm) }
\bigskip
\bigskip
\caption {\label{fig8} $m_{h}$ vs. $m_{\tilde{\chi}_{1}^{0}}$ for
$\tan{\beta}=30$ when $\sigma_{\tilde{\chi}_{1}^{0}-p}$ takes on its maximum
value.}
\end{figure}
\bigskip
\bigskip
\bigskip

\begin{figure}[htb]
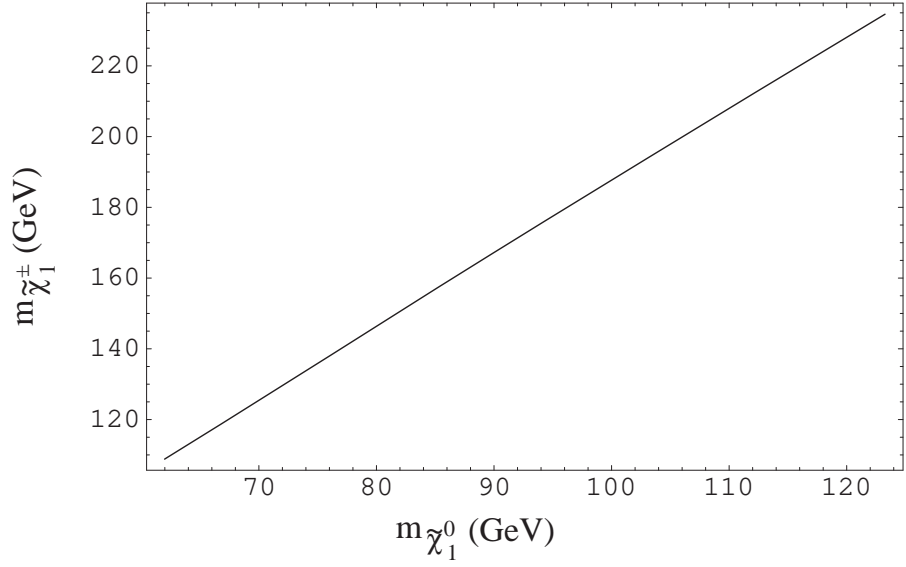

\centerline{ \DESepsf(aads30char.epsf width 12 cm) }
\bigskip
\bigskip
\caption {\label{fig9} $m_{\tilde{\chi}_{1}^{\pm}}$
vs. $m_{\tilde{\chi}_{1}^{0}}$ for $\tan{\beta}=30$ when
$\sigma_{\tilde{\chi}_{1}^{0}-p}$ takes on its maximum value.}
\end{figure}
\bigskip
\bigskip
\bigskip
\begin{figure}[htb]
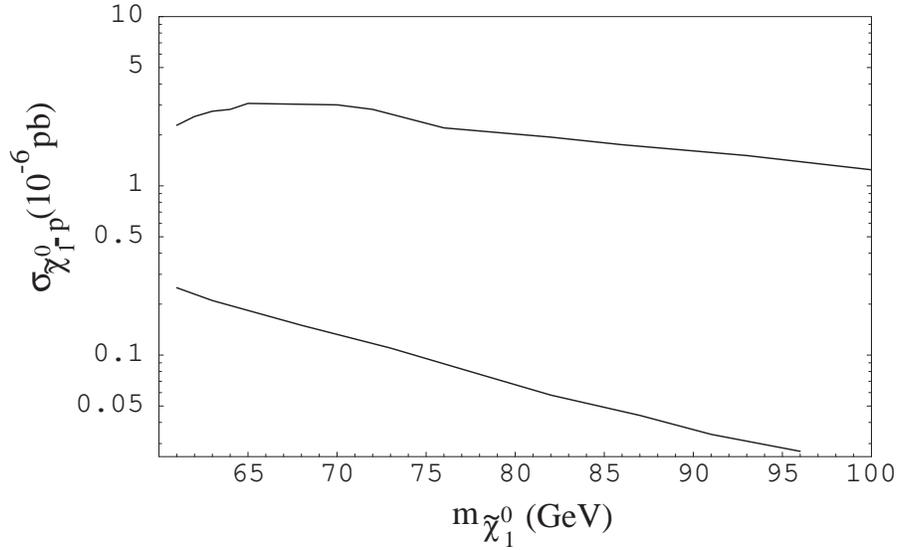

\centerline{ \DESepsf(aads7uni7nonuni.epsf width 12 cm) }
\bigskip
\bigskip
\caption {\label{fig10} Maximum value of $\sigma_{\tilde{\chi}_{1}^{0}-p}$ vs.
$m_{\tilde{\chi}_{1}^{0}}$ for $\tan{\beta}=7$ for nonuniversal soft breaking
(upper curve) and universal soft breaking (lower curve).}
\end{figure}
\bigskip
\bigskip
\bigskip

\begin{figure}[htb]
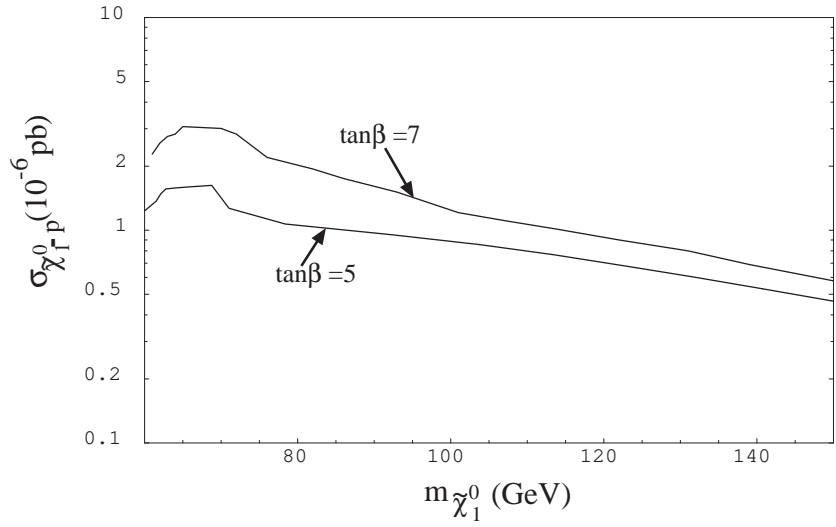

\centerline{ \DESepsf(aads357nonuni.epsf width 12 cm) }
\bigskip
\bigskip
\caption {\label{fig11} Maximum value of $\sigma_{\tilde{\chi}_{1}^{0}-p}$ vs.
$m_{\tilde{\chi}_{1}^{0}}$ for a. $\tan{\beta}=3$, and b. $\tan{\beta}=5$, 7
for nonuniversal soft breaking. Note that the $\tan{\beta}=3$ curve terminates
at $m_{\tilde{\chi}_{1}^{0}}\stackrel{\sim}{=}$ 70 GeV.}
\end{figure}
\bigskip
\bigskip
\bigskip
\begin{figure}[htb]
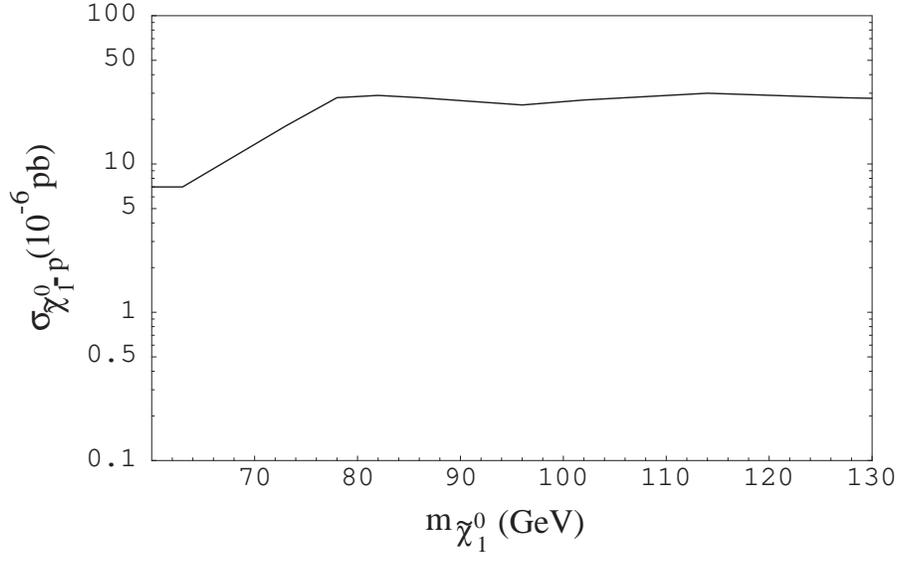

\centerline{ \DESepsf(aads15nonuni.epsf width 12 cm) }
\bigskip
\bigskip
\caption {\label{fig12} Maximum value of $\sigma_{\tilde{\chi}_{1}^{0}-p}$ vs.
$m_{\tilde{\chi}_{1}^{0}}$ for $\tan{\beta}=15$ for nonuniversal soft breaking.}
\end{figure}
\bigskip
\bigskip
\bigskip

\begin{figure}[htb]
\centerline{ \DESepsf(aads7om.epsf width 12 cm) }
\bigskip
\bigskip
\caption {\label{fig13} $\Omega_{\tilde{\chi}_{1}^{0}}h^{2}$ vs.
$m_{\tilde{\chi}_{1}^{0}}$ for maximum $\sigma_{\tilde{\chi}_{1}^{0}-p}$ for
$\tan{\beta}=7$, nonuniversal soft breaking.}
\end{figure}
\bigskip
\bigskip
\bigskip
\begin{figure}[htb]
\centerline{ \DESepsf(aads7Higgs.epsf width 12 cm) }
\bigskip
\bigskip
\caption {\label{fig14} $m_{H}$ vs.
$m_{\tilde{\chi}_{1}^{0}}$ for maximum $\sigma_{\tilde{\chi}_{1}^{0}-p}$ for
$\tan{\beta}=7$, nonuniversal soft breaking.}
\end{figure}
\bigskip
\bigskip
\bigskip

\begin{figure}[htb]
\centerline{ \DESepsf(aads7higgs1.epsf width 12 cm) }
\bigskip
\bigskip
\caption {\label{fig15} $m_{h}$ vs.
$m_{\tilde{\chi}_{1}^{0}}$ for maximum $\sigma_{\tilde{\chi}_{1}^{0}-p}$ for
$\tan{\beta}=7$, nonuniversal soft breaking.}
\end{figure}
\bigskip
\bigskip
\bigskip
\begin{figure}[htb]
\centerline{ \DESepsf(aads7dsqrk.epsf width 12 cm) }
\bigskip
\bigskip
\caption {\label{fig16} $m_{\tilde{d}}$ vs.
$m_{\tilde{\chi}_{1}^{0}}$ for maximum $\sigma_{\tilde{\chi}_{1}^{0}-p}$ for
$\tan{\beta}=7$, nonuniversal soft breaking.}
\end{figure}

\bigskip
\bigskip
\bigskip
\begin{figure}[htb]
\centerline{ \DESepsf(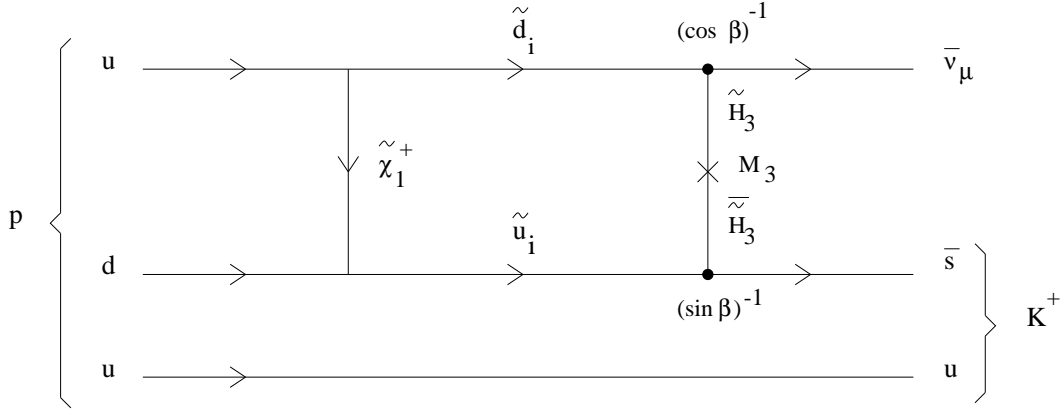 width 14 cm) }
\bigskip
\bigskip
\bigskip
\caption {\label{fig17} Example of $p$-decay diagram. The major contribution
comes from the second generation loop (i=2), the third generation contributing
$\approx 30 \%$ correction with arbitrary relative phase.}
\end{figure}

\bigskip
\bigskip
\bigskip
\begin{figure}[htb]
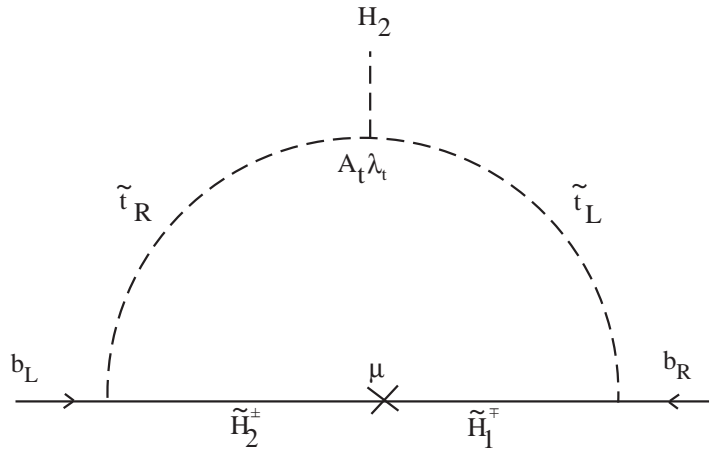

\centerline{ \DESepsf(aadsbmass.epsf width 10 cm) }
\caption {\label{fig18} One loop correction to $b$-mass.}
\end{figure}


\begin{thebibliography}{99}
\bibitem{chamseddine}A.H. Chamseddine, R. Arnowitt and P. Nath, Phys. Rev. Lett.
\textbf{49}, 970 (1982). For reviews see P. Nath, R. Arnowitt and A.H.
Chamseddine, ``Applied N=1 Supergravity" (World Scientific, Singapore, 1984);
H.P. Nilles, Phys. Rep. \textbf{110}, 1 (1984); R. Arnowitt and P. Nath, Proc.
of VII J.A. Swieca Summer School (World Scientific, Singapore, 1994).
\bibitem{arnowitt1}R. Arnowitt and P. Nath, Phys. Rev. \textbf{D54}, 2374 (1996). 
\bibitem{falk}T. Falk, K.A. Olive and M. Srednicki, Phys. Lett. \textbf{B339}, 248
(1994).
\bibitem{goldberg} H. Goldberg, Phys. Rev. Lett. \textbf{50}, 1419 (1983); J. Ellis,
J.S. Hagelin, D.V. Nanopoulos, K. Olive and M. Srednicki, Nucl. Phys.
\textbf{B238}, 453 (1984).
\bibitem{goodman}M. Goodman and E. Witten, Phys. Rev. \textbf{D31}, 3059 (1985).
\bibitem{drees}M. Drees and N. Nojiri, Phys. Rev. \textbf{D48}, 3483 (1993).
\bibitem{bottino1}A. Bottino, F. Donato, N. Fornengo and S. Scopel, Phys. Rev.
\textbf{D59}, 095003 (1999).
\bibitem{brhlik1}M. Brhlik and L. Roszkowski, hep-ph/9903468.
\bibitem{belli}P. Belli et.al., hep-ph/9903501.
\bibitem{bottino2}A. Bottino, F. Donato, N. Fornengo and S. Scopel, hep-ph/9909228.
\bibitem{kelley}S. Kelley, J. Lopez, D. Nanopoulos, H. Pois and K. Yuan, Phys. Rev.
\textbf{D47}, 246 (1993).
\bibitem{arnowitt2}R. Arnowitt and P. Nath, Phys. Rev. Lett. \textbf{70}, 3696 (1994).
\bibitem{kane}G. Kane, C. Kolda, L. Roszkowski and J. Wells, Phys. Rev.
\textbf{D49}, 6173 (1994).
\bibitem{baer1}H. Baer and M. Brhlik, Phys. Rev. \textbf{D53}, 597 (1996).
\bibitem{arnowitt3}R. Arnowitt and P. Nath, Phys. Rev. \textbf{D54}, 2374 (1996).
\bibitem{baer2}H. Baer and M. Brhlik, Phys. Rev. \textbf{D55}, 3201 (1997).
\bibitem{baer3}H. Baer and M. Brhlik, Phys. Rev. \textbf{D57}, 567 (1998).
\bibitem{barger1}V. Barger and C. Kao, Phys. Rev. \textbf{D57}, 3131 (1998).
\bibitem{ellis1}J. Ellis, T. Falk, K. Olive and M. Srednicki, hep-ph/9905481.
\bibitem{berezinsky1}V. Berezinsky, A. Bottino, J. Ellis, N. Fornengo, G. Mignola and S.
Scopel, Astropart. Phys. \textbf{5}, 1 (1996).
\bibitem{berezinsky2}V. Berezinsky, A. Bottino, J. Ellis, N. Fornengo, G. Mignola and S.
Scopel, Astropart. Phys. \textbf{6}, 333 (1996).
\bibitem{nath1}P. Nath and R. Arnowitt, Phys. Rev. \textbf{D56}, 2820 (1997).
\bibitem{arnowitt4}R. Arnowitt and P. Nath, Phys. Lett. \textbf{B437}, 344 (1998).
\bibitem{bottino3}A. Bottino, F. Donato, N. Fornengo and S. Scopel, Phys. Rev.
\textbf{D59}, 095004 (1999).
\bibitem{arnowitt5}R. Arnowitt and P. Nath, Phys. Rev. \textbf{D60}, 044002 (1999).
\bibitem{bernabei}R. Bernabei et al., Phys. Lett. \textbf{B424} (1998) 195; Phys.
Lett. \textbf{B450} (1999) 448.
\bibitem{barger2}See e.g. V. Barger, M.S. Berger and P. Ohmann, Phys. Rev.
\textbf{D49}, 4908 (1994).
\bibitem{brhlik2}M. Brhlik, L. Everett, G. Kane, and J. Lykken, hep-ph/9908326.
\bibitem{accomando}E. Accomando, R. Arnowitt and B. Dutta, hep-ph/9909333.
\bibitem{anlauf}H. Anlauf, Nucl. Phys. \textbf{B430}, 245 (1994).
\bibitem{PDG}Particle Data Group, European Physical Journal, \textbf{C3}, 1
(1998).
\bibitem{superK}Super Kamiokande Collaboration, Phys. Rev. Lett. \textbf{83}, 1529
(1999).
\bibitem{freedman}W. Freedman, astro-ph/9909076.
\bibitem{dodelson}S. Dodelson and L. Knox, astro-ph/9909454.
\bibitem{mohr}J. Mohr, B. Mathiesen and A.E. Evrad, Astrophys. J. \textbf{517},
627 (1999).
\bibitem{perlmutter}S. Perlmutter et.al., Astrophys. J. (in press) (astro-ph/9812133);
A.G. Riess et.al., Astron. J. \textbf{116}, 1009 (1998).
\bibitem{lineweaver}C. Lineweaver, astro-ph/9909301.
\bibitem{leutwyler}H. Leutwyler, Phys. Lett. \textbf{B374}, 163 (1996).
\bibitem{ellis2}J. Ellis and R. Flores, Phys. Lett. \textbf{B263}, 259 (1991);
\textbf{B300}, 175 (1993).
\bibitem{jungman}For reviews see G. Jungman, M. Kamionkowski and K. Greist, Phys. Rep.
\textbf{267}, 195 (1995); E.W. Kolb and M.S. Turner, ``The Early Universe" (Addison-Wesley,
Reading, 1990).
 
\bibitem{greist}K. Greist and D. Seckel, Phys. Rev. \textbf{D43}, 3191 (1991); P
Gondolo and G. Gelmini, Nucl. Phys. \textbf{B360}, 145 (1991).
\bibitem{arnowitt6}R. Arnowitt and P. Nath, Phys. Lett. \textbf{B299}, 58 (1993);
Erratum ibid \textbf{B303}, 403 (1993); P. Nath and R. Arnowitt, Phys. Rev.
Lett. \textbf{70}, 3696 (1993).
\bibitem{roszkowski}L. Roszkowski, Phys. Lett. \textbf{B262}, 59 (1991).
\bibitem{arnowitt7}R. Arnowitt and P. Nath, Phys. Rev. Lett. \textbf{69}, 725 (1992).
\bibitem{goto}T. Goto and T. Nihei, Phys. Rev. \textbf{D59}, 115009 (1999).
\bibitem{rattazi}R. Rattazi and U. Sarid, Phys. Rev. \textbf{D53}, 1553, (1996); M.
Carena, M. Olechowski, S. Pokorski and C. Wagner, Nucl. Phys.
\textbf{B426}, 269, (1994).
\end{thebibliography}
 \end{document}